\documentclass[a4paper,12pt]{article}
\pdfoutput=1
\usepackage{jcappub}
\usepackage{amsthm}
\usepackage{graphicx}
\usepackage{color}
\usepackage{graphicx,textcomp,float,gensymb,wrapfig, enumitem,comment,dsfont,subfigure,framed,slashed,appendix,wrapfig}
 
 \newcommand{\be}{\begin{equation}}
\newcommand{\ee}{\end{equation}}
\newcommand{\MP}{M_{\rm P}}

\newcommand{\hc}{+\,\mathrm{h.c.}}
\newcommand{\SU}[1]{\ensuremath{\mathrm{SU}(#1)}}
\newcommand{\U}[1]{\ensuremath{\mathrm{U}(#1)}}
\newcommand{\tr}{\operatorname{tr}}
\newcommand{\into}{\ensuremath{\,\rightarrow\,}}
\def\bsp#1\esp{\begin{split}#1\end{split}}
\def\bpm{\begin{pmatrix}}
\def\epm{\end{pmatrix}}

\newcommand{\bea}{\begin{eqnarray}}  
\newcommand{\eea}{\end{eqnarray}}  
 \def\bsp#1\esp{\begin{split}#1\end{split}}
\title{GUT-scale inflation with sizeable tensor modes}

\author[a]{Felix Br\"ummer,}
\author[a]{Valerie Domcke}
\author[b]{and Ver\'onica Sanz}

 \affiliation{
$^a$ SISSA/INFN, Via Bonomea 265, 34136 Trieste, Italy
}
 \affiliation{
$^b$ Department of Physics and Astronomy, University of Sussex, Brighton BN1 9QH, UK
}

\abstract{A sizeable tensor-to-scalar ratio, such as recently claimed by BICEP2, would imply a scale of inflation at the typical scale of supersymmetric grand unification. This could be an accident, or strong support for supersymmetric theories. Models of F-term hybrid inflation naturally connect the GUT scale with the inflationary scale, but they also predict the tensor-to-scalar ratio to be unmeasurably small. In this work we analyze a general UV embedding of F-term hybrid inflation into a supergravity theory with a general K\"ahler potential. The CMB observables are generated during the early phase of inflation, at large inflaton values, where the potential is dominated by Planck-suppressed operators. Tuning the leading higher-order terms can give an inflaton potential with sizeable tensor fluctuations and a field excursion which is still sub-Planckian but close to the Planck scale, as expected from the Lyth bound.
}

\emailAdd{fbruemmer@sissa.it}
\emailAdd{vdomcke@sissa.it}
\emailAdd{v.sanz@sussex.ac.uk}


\begin{document}
\maketitle
\flushbottom

\section{Introduction}

Recently the BICEP2 collaboration reported a discovery of B-mode polarization in the cosmic microwave background~\cite{Ade:2014xna}. When interpreted as originating from primordial gravitational waves from inflation, this signal corresponds to a tensor-to-scalar ratio of around $r \sim 0.2$, and to a scale of inflation around $M_{\rm GUT}=2 \times 10^{16}$ GeV. 
This claim is currently under intense scrutiny; however, even if the amplitude of the primordial tensor fluctuations were an order of magnitude smaller than reported, this
would still point towards a scale of inflation near the scale of supersymmetric grand unification.
Hence, a detection of primordial gravity waves immediately has two remarkable consequences: First, it suggests a connection between inflation and SUSY GUTs. Second, assuming single-field slow-roll inflation with a potential whose slope increases monotonically until slow-roll is violated at the end of inflation, it implies super-Planckian field values during inflation~\cite{Boubekeur:2005zm}, $\sigma \geq N \sqrt{r/8} \MP \sim 8 \MP$.

These two implications are difficult to reconcile. On the one hand, at such large field values, the theory is generically dominated by uncontrolled higher-dimensional operators.
They can be forbidden by symmetries, but these symmetries, against generic expectations, would need to be respected by quantum gravity. Hence inflation might be governed by physics which is not described by supergravity as an effective field theory, and the appearance of the GUT scale might be a pure coincidence. On the other hand, taking the hint for linking inflation and supersymmetric grand unification seriously typically leads to small-field models. Inflation is linked to a GUT-breaking phase transition, with field values $M_{\rm GUT} < \sigma \ll \MP$ rendering these models calculable in an effective supergravity framework. A prototypical example is $F$-term hybrid inflation (FHI) \cite{Copeland:1994vg, Dvali:1994ms}: The inflaton potential is essentially logarithmic, which allows to reach 50 - 60 e-folds within a small field 
range but at the same time predicts $r$ to be very small. 

In this paper, we aim to find a compromise between these two seemingly conflicting observations. We take the connection of inflation and supersymmetric grand unification seriously, by studying a variant of FHI which is neither strictly small-field nor strictly large-field inflation. When allowing the inflaton $\sigma$ to reach values close to (but not exceeding) the Planck scale, $\MP$-suppressed operators become important. As was shown in \cite{XV} for general models of inflation, these may significantly affect the shape of the inflaton potential and thus the prediction for $r$. Relaxing all assumptions on the monotonicity of the inflaton potential allows us to realize $r\sim{\cal O}(0.1)$ with sub-Planckian field values, avoiding the bound on the field excursion of \cite{Boubekeur:2005zm} but in accordance with the weaker bound of \cite{Lyth:1996im} and the studies of \cite{BenDayan:2009kv, Hotchkiss:2011gz, Antusch:2014cpa}. See also Ref.~\cite{Hebecker:2013zda} for related work in the context of natural 
inflation. Simultaneously, the Planck-
suppressed operators can account for a sufficiently red-tilted spectral index, $n_s = 0.96$, cf.\ \cite{BasteroGil:2006cm}.

In the context of FHI we can obtain suitable potentials by allowing for $R$-symmetry breaking terms in the K\"ahler potential, suppressed by powers of $\MP$.\footnote{The effects of Planck-suppressed operators from $R$-symmetry preserving non-canonical K\"ahler terms were recently studied in \cite{Civiletti:2014bca}. However, the result of this study was that $r$ remains small even when taking those terms into account: generally $r\lesssim 0.01$ for sub-Planckian inflaton values.} This is in line with the expectation that the continuous global $R$-symmetry which governs the superpotential of ordinary FHI will be broken by quantum gravity effects. The inflaton potential reduces to that of FHI in the global SUSY limit $\MP\into\infty$. During the early inflationary phase which determines the cosmic microwave background (CMB) observables, it is however completely dominated by higher-dimensional operators (whose coefficients must be suitably tuned to obtain the desired values for $r$ and $n_s$ without 
violating the Planck bounds on the 
running, and the running of the running, of the spectral index).
We find that, working with terms up to the order $1/\MP^6$, our model can saturate the Lyth bound \cite{Lyth:1996im}, leading to inflaton values  $\sigma\gtrsim 0.4\,\MP$ for $r\approx 0.1$. 
For such large values of $r$, terms of even higher order are therefore generically not under good control. Our conclusion is that this model can be reconciled with the Planck and BICEP2 data, but only by judiciously choosing the parameters entering the scalar potential up to rather high order in the $1/\MP$ expansion.

\section{{\it F}-term hybrid inflation and its supergravity embedding}

Recall that the superpotential in $F$-term hybrid inflation is linear in the inflaton superfield $S$,
\be
W=\lambda\, S\left(\Lambda^2-Q\tilde Q\right)\,.
\ee
There is a global continuous $\U{1}_R$ symmetry under which $S$ carries charge $2$ and $Q$, $\tilde Q$ are neutral. During the inflationary phase, the inflaton superpotential reduces to
\be\label{WFHIglobal}
W_{\rm FHI}=\lambda \Lambda^2\,S\,,
\ee
where $\sqrt{\lambda}\Lambda$ is the scale of inflation.  With a canonical K\"ahler potential, and in the limit $\MP\into\infty$, the scalar potential is exactly flat at the tree level. At the one-loop level, there is a logarithmic correction coming from $Q$ and $\tilde Q$, which owe their mass to the inflaton,
\be\label{VCW}
V_{\text{CW}}(S)=\frac{C\,\lambda^4}{16\pi^2}\Lambda^4\log\left(\frac{\lambda^2|S|^2}{\mu^2}\right)\,.
\ee
Here $C$ is a group-theoretical prefactor and $\mu$ is the renormalization scale. The tensor-to-scalar ratio is small in this model, even when incorporating higher-order terms from supergravity and from a non-canonical K\"ahler potential respecting $\U{1}_R$ \cite{Civiletti:2014bca}.

The $R$-symmetry must, at the latest, be broken at some point after inflation in order to obtain an (almost) Minkowski vacuum with broken supersymmetry. Depending on the model, it may also be broken explicitly by renormalizable inflaton couplings to other fields, as in the model of \cite{ddr} which we review in Appendix~\ref{app_ddr}. In any case we do not expect quantum gravity effects to respect global symmetries, so it is reasonable to supplement the superpotential Eq.~\eqref{WFHIglobal} by a K\"ahler potential of the form
\be\label{KFHI}
K=|S|^2+\sum_{m+n\,\geq\, 3}\left( k_{mn}\frac{S^m(S^\dag)^n}{\MP^{m+n-2}}\hc\right) \,,
\ee
where we have assumed $S$ to be canonically normalized up to quadratic order. We also allow for a constant term in the superpotential,
\be\label{WFHI}
W=W_0+\lambda\Lambda^2\,S\,.
\ee
 Note the absence of terms such as $\Lambda\, S^2$ or $S^3$ in $W$. This structure could be the result of the $R$-symmetry being broken in a separate sector with vanishing or at most very small couplings to the inflaton sector. Gravitational physics will still communicate $R$-breaking to the inflaton, but only in the form of $\MP$-suppressed operators in $K$ as in Eq.~\eqref{KFHI}.  (Of course, some of these operators can partly be absorbed in $W$ by a K\"ahler-Weyl transformation $K\into K+f+f^\dag$, $W\into W e^{-f/\MP^2}$, but the resulting corrections to $W$ are always suppressed by powers of $\MP$ and thus very small, as opposed to, say, a $\Lambda\,S^2$ inflaton mass term which would completely upset the model.)

Since all $R$-symmetry breaking corrections are suppressed by powers of $\MP$, we recover FHI in the rigid limit. On the other hand, for inflaton values of ${\cal O}(\MP)$ the coefficients $k_{mn}$ and the free parameter\footnote{It is tempting to identify $W_0/(3\MP^2)$ with the gravitino mass as expected for TeV-scale supersymmetry, and to therefore impose $W_0\lesssim (10^{13}\,\text{GeV})^3$. Here we will make no such assumption, since SUSY may as well be broken at a higher scale, or there may be other contributions to the gravitino mass after the end of inflation. For instance, in the model of \cite{ddr}, $W_
0$ must be of the order of $(M_{\rm GUT})^3$ to cancel the cosmological constant after inflation has ended (see Appendix~\ref{app_ddr} for details). We will however assume $W_0/\MP\ll\lambda\Lambda^2$ throughout this paper.} 
$W_0$ can take values which allow the CMB observables to significantly deviate from the FHI predictions, as we will detail in the next Section.

The scalar potential is
\be\label{VS}
V(S)=e^{K/\MP^2}\left(|D_S W|^2 K^{\bar S S}-3\frac{|W|^2}{\MP^2}\right)+V_{\text{CW}}(S) \,,
\ee
where we approximate $V_{\text{CW}}$ by the one-loop Coleman-Weinberg potential in the globally supersymmetric limit, cf.~Eq.~\eqref{VCW}, discarding terms that are doubly suppressed by both $\MP$ and by a loop factor. 

The $R$-symmetry breaking terms in the superpotential and in the K\"ahler potential break the degeneracy in the phase of the complex inflaton field $S$, turning FHI into a two-field inflation model~\cite{Buchmuller:2014epa}. However, the real axis remains a self-consistent solution provided that $W_0$ and the $k_{ij}=k_{ji}$ are real, and in order to prove that this setup can account for a large tensor-to-scalar ratio, it will be sufficient to focus on this solution. We leave the analysis of the full two-field model to future work.
The canonically normalized inflaton field is now
\be
\sigma=\frac{1}{\sqrt{2}}\; {\rm Re}\;S\,,
\ee
and its scalar potential can be written as 
\be\label{Vsigma2}
V(\sigma)=V_0-3\frac{W_0^2}{\MP^2}+\frac{C\lambda^2}{16\pi^2}V_0\log\frac{\lambda^2\sigma^2}{2\mu^2}+V_0\sum_{n\geq 1}a_n\frac{\sigma^n}{\MP^n} \,,
\ee
where the coefficients $a_n$ are function of the parameters $k_{ij}$ in the K\"ahler potential, and we have defined
\be
V_0=\lambda^2\Lambda^4\,.
\label{eq_V0}
\ee
The expressions for the coefficients $a_n$ of the higher-order terms quickly become very unwieldy in this expansion. For practical purposes it is more convenient to work directly in terms of the coefficients $a_n$ appearing in the scalar potential,
as we will do in the following. Any potential in which the $\MP$-suppressed terms take the general form of Eq.~\eqref{Vsigma2} can be obtained by choosing $k_{ij}$ and $W_0$ suitably. In Appendix~\ref{app_coeff}, we give a translation between the terms in the K\"ahler and the leading coefficients in Eq.~\eqref{Vsigma2}.

\section{An upper bound on the tensor-to-scalar ratio
\label{sec_inflation}}

Our goal is now to find a parameter region in which the potential Eq.~\eqref{Vsigma2} gives rise to inflation with a relatively large tensor-to-scalar ratio and a relatively small field excursion $\Delta \sigma$, such that higher-order terms in the $\sigma/\MP$ expansion are under control. 
The free parameters entering Eq.~\eqref{Vsigma2} are the global SUSY vacuum energy during inflation $V_0$ defined by Eq.~\eqref{eq_V0},
the superpotential coupling constant $\lambda$, the constant term in the superpotential $W_0$, and the coefficients of the last term in Eq.~\eqref{Vsigma2}, which we shall truncate at ${\cal O}(1/\MP^6)$ leaving us with the parameters $a_{1,2 \dots 6}$. Moreover, in the following we set $C = 1$ (a different value can always be absorbed into a redefinition of $\lambda$ and $\mu$) and $\mu = \lambda \Lambda$ (our results are not sensitive to the precise choice of the renormalization scale).

 As it will turn out, truncating the series at the order $(\sigma^6/\MP^6)$ does not mean that higher terms can safely be neglected for generic ${\cal O}(1)$ values of the coefficients $a_{\geq 7}$. By definition it is clear that, for any small-field model, eventually higher terms become negligible as long as the sequence of the $\{a_n\}$ is well-behaved, but in our case (as should become clear below) this point is only reached at much higher order. In fact we will need to fine-tune the leading ${\cal O}(10)$ coefficients, and setting $a_{\geq 7}$ to zero merely corresponds to a specific choice for this fine-tuning. 

To determine a viable parameter region, observe that the slow-roll parameters $\epsilon$ and $\eta$, defined by
\be
\epsilon=\frac{\MP^2}{2}\left(\frac{V'}{V}\right)^2\,,\qquad\qquad\eta=\MP^2\frac{V''}{V}\,,
\ee
are now fixed at the CMB pivot scale by observation. That is,
using the values published by the BICEP2~\cite{Ade:2014xna} and Planck~\cite{Ade:2013uln} for $r$ and $n_s$ respectively, we have
\be
r=16\,\epsilon_* = 0.2^{+0.07}_{-0.05}\,,\qquad\qquad n_s=1-6\,\epsilon_*+2\,\eta_* =  0.9600 \pm 0.0071\,,
\label{eq_obs1}
\ee
where $\epsilon_*$ and $\eta_*$ refer to the slow-roll parameters evaluated at $\sigma = \sigma_*$, $N_* = 50 - 60$ e-folds before the end of inflation. It should be noted that the value of $r$ 
may decrease depending on the foreground dust model which one subtracts,
$r = 0.16^{+0.06}_{-0.05}$~\cite{Ade:2014xna}, bringing it into somewhat better agreement with the upper bound $r < 0.11$ from Planck~\cite{Ade:2013uln}. Future measurements of the B-mode spectrum and possible dust foregrounds will be crucial for a precise determination of $r$.

From now on we assume that the vacuum energy during inflation is dominated by the global SUSY term $V_0$, i.e.~that the second term on the RHS of Eq.~\eqref{Vsigma2} can be neglected.
Then $V_0$ is fixed by the amplitude of the power spectrum~\cite{Ade:2013uln},
\be
A_s = \frac{V(\sigma_*)}{24 \pi^2 \epsilon_*} = 2.20^{+0.05}_{-0.06} \times 10^{-9}\,.
\label{eq_As}
\ee
Given the structure of the scalar potential Eq.~\eqref{Vsigma2}, the parameter $V_0$ enters only into $A_s$, but cancels in all the slow-roll parameters and in the slow-roll equation of motion. 
Thus, after ensuring the correct slow-roll dynamics, we can always determine $V_0$ a posteriori using Eq.~\eqref{eq_As}.

We can further restrict the parameter space analytically by confronting our model with the Lyth bound \cite{Lyth:1996im}. With no assumptions on the monotonicity properties of the inflaton potential, except that $\epsilon$ should be approximately constant during the first $N_0\sim 4-5$ e-folds which leave their imprint on the CMB observables, the field excursion is bounded from below as (see also \cite{Lyth:2014yya})
\be\label{eq_lyth_N5}
\Delta\sigma\gtrsim N_0\sqrt\frac{r}{8}\,\MP\, \Rightarrow \frac{\sigma_*}{\MP} \simeq 0.45 \,  \sqrt{\frac{r}{0.1}} \ . 
\ee
If we want to succeed in reproducing a large tensor-to-scalar ratio for $\Delta\sigma <\MP$, we must get at least close to saturating Eq.~\eqref{eq_lyth_N5}. Hence we need a scalar potential which is rather steep for the first ${\cal O} (5)$ e-folds and then quickly becomes very flat to accommodate the remaining $\sim 50$ e-folds. Achieving this with an analytic single-field inflation potential as in Eq.~\eqref{Vsigma2} implies that the higher-order derivatives of $V$ are generically large (see also \cite{BenDayan:2009kv}), typically $|V^{(n)}/V| \sim {\cal O}(1)$, much larger than preferred by the Planck data~\cite{Ade:2013uln}. However, since the sign of $V^{(n)}$ is not fixed, $|V^{(n)}/V|$ will have zeros in which the coefficients in Eq.~\eqref{Vsigma2} conspire so that the higher derivatives are small. Hence without loss of generality we can set the higher-dimensional slow-roll parameters at horizon crossing to an arbitrary value allowed by the Planck data,\footnote{Note that the bound on 
$\alpha_s$ depends on the inclusion of a non-vanishing $r$ and $\kappa_s$. With no bounds given by the Planck collaboration including both additional parameters, we here opt for using the fit including $\kappa_s$. We expect that including the non-vanishing $r$ should not significantly alter the best-fit value, but will possibly enlarge the error bands.}
thus eliminating two further parameters~\cite{Liddle:1994dx, Ade:2013uln}:
\begin{equation}
 \alpha_s \simeq - 2 \frac{V' V^{(3)}}{V^2} \stackrel{!}{=} 0.001^{+ 0.013}_{-0.014}\,, \quad \kappa_s \simeq  2 \frac{V'^2 V^{(4)}}{V^3} \stackrel{!}{=} 0.022^{+0.016}_{-0.013} \,.
\label{eq_obs3}
\end{equation}
As expected, varying these bounds within the experimental errors does not change the qualitative picture, but does impact the quantitative results somewhat. We find that in order to achieve our goal of large $r$ for moderate $\sigma_*$, the most convenient choice is to set both $\alpha_s$ and $\kappa_s$ to their experimental upper bound.
This is in agreement with analyses based on flow equations~\cite{Hansen:2001eu, Caprini:2002jy}, which indicate that a very flat potential at the end of inflation can be achieved for sufficiently large values of $\alpha_s$ and $\kappa_s$.

Finally, again exploiting the Lyth bound, we know that the initial value $\sigma_*$ of the inflaton field will need to be sizeable for the $\MP$-suppressed operators to be relevant, but not too large in order to retain control over the subdominant terms. Fixing $\sigma_*={\cal O}(0.3 -0.6)\,\MP$ and using Eqs.~\eqref{eq_obs1} to \eqref{eq_obs3} to eliminate five of the eight remaining parameters in Eq.~\eqref{Vsigma2} allows us to restrict the final three free parameters, subject to the consistency condition that $N_*=50$ e-folds are realized.

\begin{figure}
\centering
 \includegraphics[width = 0.7\textwidth]{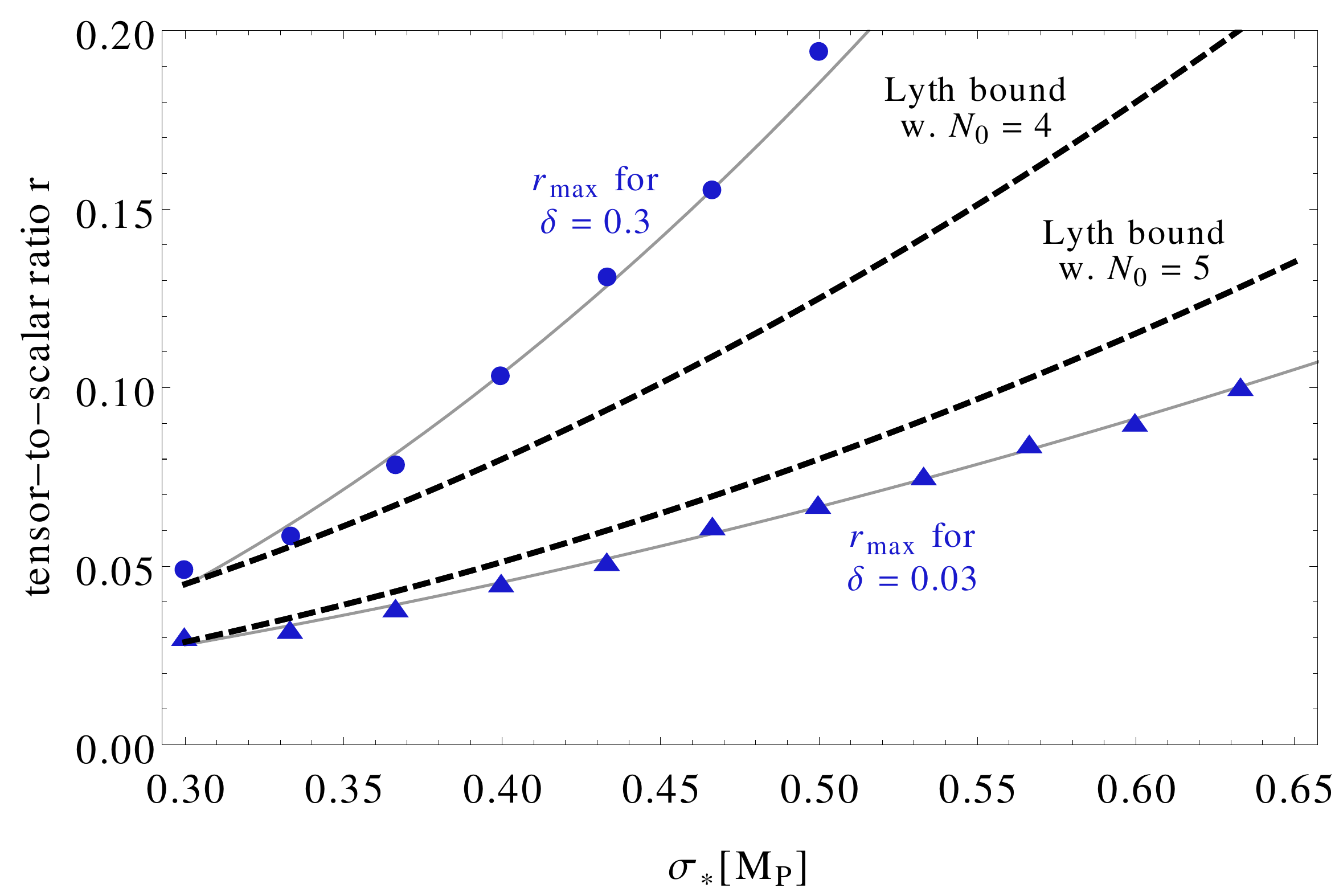}
\caption{Maximal tensor-to-scalar ratio as a function of $\sigma_*$. We require $N_* = 50$ e-folds, the values of $n_s$ and $A_s$ to be at their best fit values, and $\alpha_s$ and $\kappa_s$ within the $3 \sigma$ range. We also require  all higher order derivatives to be small, in particular $\delta = 0.3$ (blue dots) and $\delta = 0.03$ (blue triangles). For comparison we also show the Lyth bound, assuming $\epsilon$ to be constant over the first 4 (5) e-folds (black-dashed lines).}
\label{fig_sigma_vs_r}
\end{figure}

Fig.~\ref{fig_sigma_vs_r} illustrates the result of this analysis,
showing the maximal tensor-to-scalar ratio achievable for a given value of $\sigma_*$ in this framework. Starting from the scalar potential in Eq.~\eqref{Vsigma2}, we require $N_* = 50$ e-folds of slow-roll inflation, $n_s$ and $A_s$ to be at their best fit values according to Eqs.~\eqref{eq_obs1} and \eqref{eq_As}, $\alpha_s$ and $\kappa_s$ to be at their respective $3 \sigma$ upper limit according to Eqs.~\eqref{eq_obs3}. We then perform a parameter scan in the remaining three free parameters, requiring all higher order derivatives to be under control, $|2 (V')^{n-2} V^{(n)}/V^{n-1}| < \delta$, with $\delta = 0.3$ and $\delta = 0.03$ serving as representative examples for a less and a more conservative bound, respectively. For comparison we also show the Lyth bound, cf.\ Eq.~\eqref{eq_lyth_N5}, with $N_0 = 4(5)$. We find that our the setup of generalized hybrid inflation discussed in this paper can indeed saturate (and slightly exceed) even the most conservative Lyth bound, 
allowing for a relatively large tensor-to-scalar ratio ${\cal{O}}(0.1)$ for a moderate value of the inflaton field ${\cal{O}}(0.5)\,\MP$, marginally justifying 
the expansion in $\MP$ suppressed operators even in the light of the BICEP2 result, at the price of tuning the coefficients of the first few operators. 

We should point out here that there are two kinds of observational constraints which might still threaten the validity of these parameter points.\footnote{We thank Shaun Hotchkiss for helpful discussions on these issues.} Firstly, a too flat potential towards the end of inflation may lead to overproduction of primordial black holes, see e.g.~\cite{Carr:2009jm} for an analysis of the resulting bounds. Secondly, the CMB data severely constrains any variation in the power spectrum amplitude over the first few e-folds around $\sigma_*$ \cite{Ade:2013zuv}, so even the more conservative assumption $\delta = 0.03$ might turn out not to be conservative enough. We have verified that, while the constraints on primordial black holes do rule out a part of the parameter space, there are nevertheless many valid parameter points left even towards the region of low $\sigma_*$ and low $r$ (and in particular, on the curves shown in Fig.~\ref{fig_sigma_vs_r}). As to the second point, addressing it would require a rather 
involved analysis beyond the slow-roll approximation and beyond the derivative expansion for the potential. A tentative check (which still partly relies on the slow-roll approximation despite $\eta$ briefly becoming ${\cal O}(1)$ in our scenario) seems to indicate that one may need to go to values of $\delta$ even smaller than $0.03$ such as to keep the variation of the power spectrum amplitude under control. However, the situation is not conclusive, and further study is needed to settle this issue.

What characterizes the scalar potentials  which lead to large values of $r$ for moderate values of $\sigma_*$? To achieve a large value of $r$, the potential must feature a rather large first derivative at $\sigma \sim \sigma_*$, when the CMB scales left the horizon, while all higher derivatives should be small to satisfy the Planck constraints. To account for a small value of $\sigma_*$, this linear behaviour of the potential must transition to a very flat part of the potential which accounts for most of the e-folds at small field values. The total field excursion $\sigma_*$ is minimized if this transition is sharp, typically rendering higher-order derivatives large. This renders our upper bound on the tensor-to-scalar ratio sensitive to the bounds imposed on the higher slow-roll parameters, cf.~Fig.~\ref{fig_sigma_vs_r}. It also implies that future measurements constraining these higher derivatives will be crucial to test this class of `intermediate-field' models.\footnote{With a sufficient amount of 
tuning in the parameters, the impact on the higher derivatives can be suppressed/delayed to the (yet unreported) value of the higher-derivatives of the inflationary potential. In this case, a direct comparison with the observed temperature two-point function might prove more restrictive than the usual expansion in $d^{n} n_s/ d\ln k^n$, see e.g.\ \cite{BenDayan:2009kv} for a related analysis.} Moreover, we note that we do not saturate the upper bound on the tensor-to-scalar ratio recently published in \cite{Antusch:2014cpa}, which might naively allow for $r \sim 0.1$ for field excursion $\Delta \sigma \sim 0.1 \MP$ under the slow-roll condition $\epsilon, \eta \lesssim 1$. The reason, as pointed out also by the authors of Ref.~\cite{Antusch:2014cpa}, are again the constraints on the higher-order derivatives ($V'''$ and beyond) which enforce the transition between the two phases (large $\epsilon$ and small $\epsilon$) to happen gradually. 

\begin{figure}
\centering
\begin{minipage}{0.48\textwidth}
  \includegraphics[width = 1\textwidth]{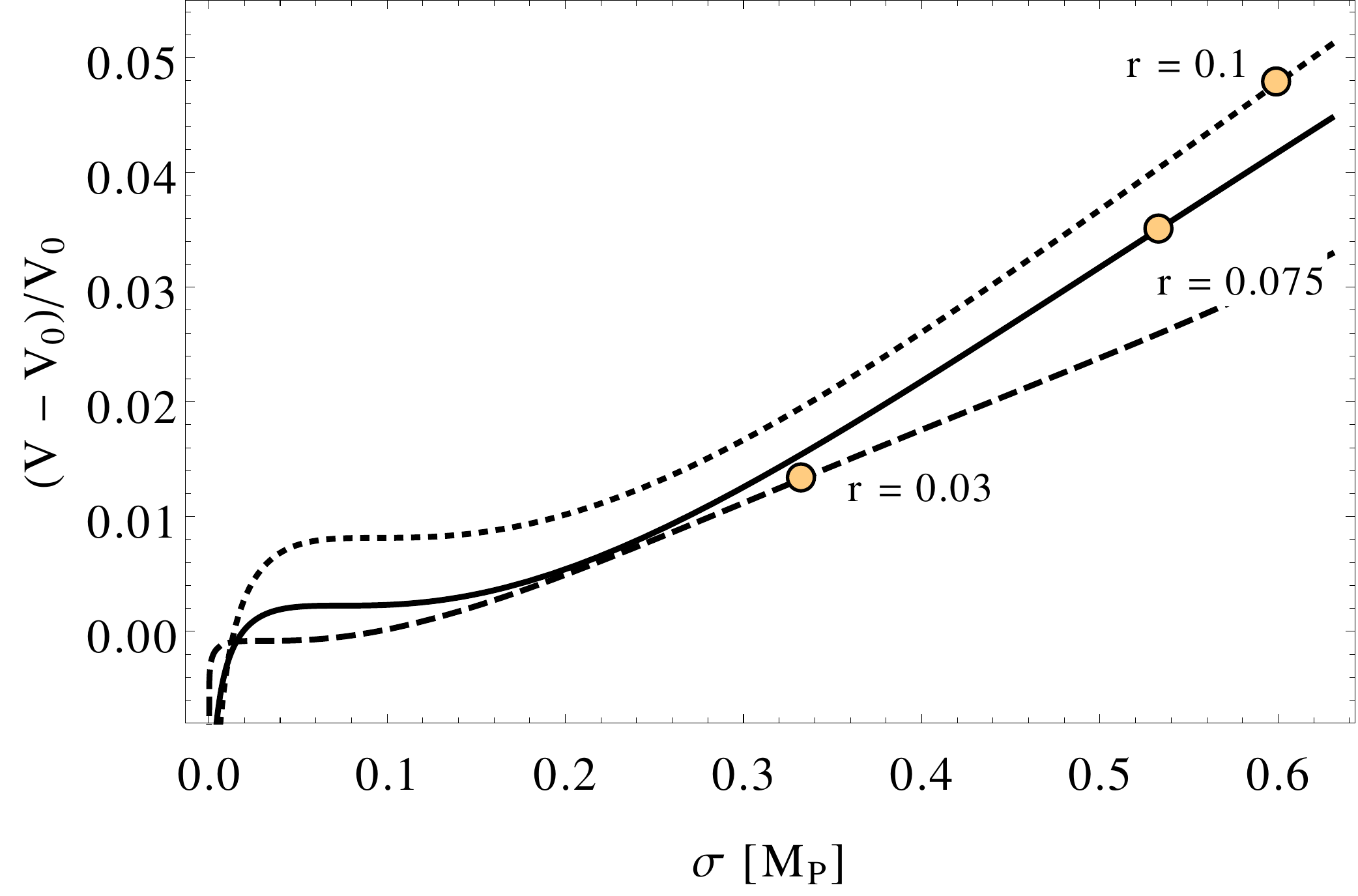}
\end{minipage} \hfill
\begin{minipage}{0.48\textwidth}
 \includegraphics[width = 1\textwidth]{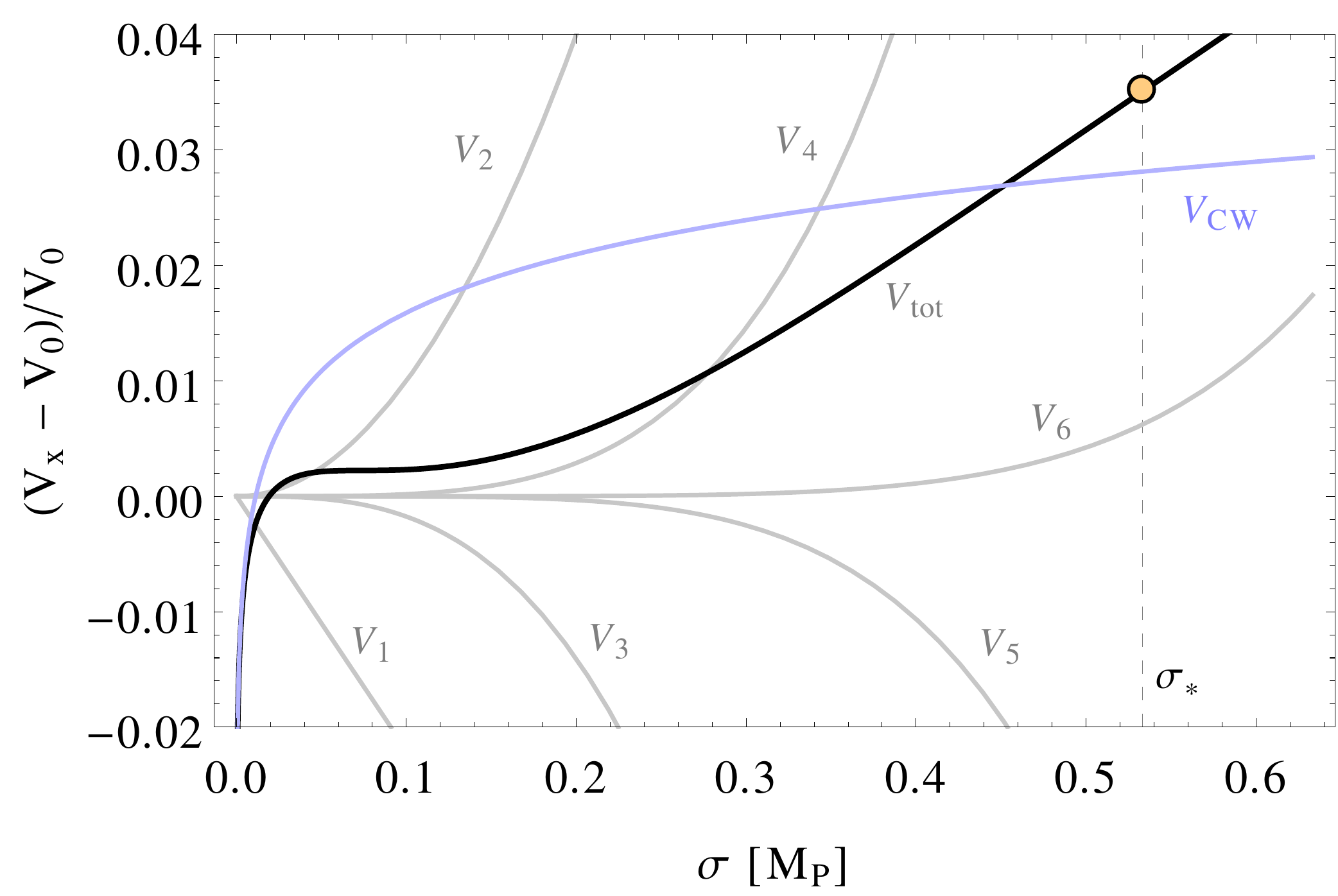}
\end{minipage}
\caption{A typical example for a scalar potential saturating the upper bound on the tensor-to-scalar ratio $r$ for the conservative case $\delta = 0.03$. Left panel: different pairs of $\sigma_*$ (orange markers) and corresponding $r$. Right panel: Decomposition of the total scalar potential for $r = 0.075$ (solid black) into the global SUSY contribution $V_{\rm CW}$ (blue) and the supergravity contributions $V_n =  a_n V_0 (\sigma/\MP)^n$ (gray). }
\label{fig_potential}
\end{figure}

Fig.~\ref{fig_potential} shows this behaviour of the scalar potential for some of the parameter combinations saturating the upper bound on $r$ in Fig.~\ref{fig_sigma_vs_r}. The left panel demonstrates how a larger tensor-to-scalar ratio impacts the linear part of the potential, leading to larger field excursion. The right panel shows the decomposition of the total scalar potential in terms of its various contributions according to Eq.~\eqref{Vsigma2}. Generically, the individual supergravity contributions are large, and as expected, it is necessary to tune the coefficients in order to obtain the desired shape of the potential. The globally supersymmetric contribution $V_{\rm CW}$ is responsible for ending inflation but is subdominant for most of the inflationary trajectory. To give an example, the parameters for the potential depicted in the right panel of Fig.~\ref{fig_potential} are $a_1 = -0.22$, $a_2 = 1.0$, $a_3 = -1.8$, $a_4 = 1.8$, $a_5 = -1.0$, $a_6 = 0.27$, $\lambda = 0.76$ and $V_0 = (1.7 \times 
10^{16}~\text{GeV})^4$. This is a quite typical example, in the sense that all dimensionless coefficients are ${\cal O}(1)$ and $V_0$ (as determined by $A_s$ and $r$, cf.\ Eq.~\eqref{eq_As}) is close to the SUSY GUT scale. Fig.~\ref{fig_potential} also suggests that, if the coefficient $a_7$ which we switched off had instead an ${\cal O}(1)$ value, the behaviour of the potential would be significantly changed at large $\sigma$. As stated above, our motivation for truncating the potential at $a_6$ is not that higher terms are negligibe (which is the case only for $a_{\gtrsim 10}$ in this example) but that nonzero $a_{1\ldots 6}$ is the minimal set of nonzero coefficients needed.

What could be the physical reason justifying this particular shape of the scalar potential? The two regimes (characterized by an approximately flat potential at small $\sigma$, and an approximately linear potential at large $\sigma$) cannot be simply identified with the domains where $\MP$-suppressed terms are respectively negligible and dominant. This can be seen from the right panel of Fig.~\ref{fig_potential}, which shows that various $\MP$-suppressed terms contribute significantly to flattening the potential at small $\sigma$, together with the Coleman-Weinberg contribution. Furthermore, the transition between these two domains is generically smooth, while to obtain a large tensor-to-scalar ratio with minimal field excursion, one would prefer them to be linked by a sharper bend. One may speculate that the sudden increase in the potential might be triggered by a phase transition, and that the linear regime at large field values might be an exact feature of the UV theory of quantum gravity (for instance, 
it might be somehow linked to the linear inflaton potentials in axion monodromy inflation in superstring theory \cite{McAllister:2008hb}). In the present work, however, we merely show that tuning the parameters of the scalar potential can mimic such effects.

\section{Conclusion and Outlook}

In this paper we have revisited models of $F$-term hybrid inflation, which are well known to link the inflationary scale with the scale of supersymmetric grand unification. In its minimal form, FHI also predicts an unobservable small tensor-to-scalar ratio. Motivated by the recent results from BICEP2, we have therefore considered a non-minimal extension of FHI  by $\MP$-suppressed operators.

In fact, if the field excursion of the inflaton is of the order of the Planck scale, the scalar potential is dominated by $\MP$-suppressed terms which can significantly change the predictions for the CMB observables. Since quantum gravity is expected to break global symmetries, these $\MP$-suppressed operators generically include non-minimal K\"ahler terms which explicitly break the $R$-symmetry governing minimal FHI. Treating the coefficients of these terms as free parameters, one can tune them to obtain an inflaton potential which is approximately linear during the first $4-5$ e-folds, and then approximately constant for the remaining $\approx 50$ e-folds. This allows for a sizeable tensor-to-scalar ratio for sub-Planckian values of the inflaton field. The potential reduces to that of FHI only at very small inflaton values, where $\MP$-suppressed terms are negligible. One thus retains the connection between the GUT scale and the inflationary scale, and more generally the connection to field-theoretical 
particle physics model building as in FHI models, while the predictions for the inflationary observables depend mostly on Planck-scale physics. 

By carefully choosing the coefficients of the higher-dimensional operators, we can obtain models where the inflaton field excursion saturates the Lyth bound. Since this bound already implies a minimal field excursion of the order of $\approx 0.5\,\MP$ for a tensor-to-scalar ratio of $r=0.1$, the K\"ahler potential must be tuned to a fairly high order in the $1/\MP$ expansion in order to reproduce the inflationary observables correctly. Nevertheless, we find it interesting that one can write down a model with (slightly) sub-Planckian field values which connects the GUT scale with the scale of inflation, and which allows for a sizeable value of $r$.

At the small field values at the end of inflation the supergravity contributions become negligible. We hence expect the subsequent cosmological processes, i.e.\ the generation of a thermal bath, of a matter-antimatter asymmetry and of dark matter to proceed as in the globally supersymmetric case, albeit with a larger value for the superpotential coupling
$\lambda$ as well as for the energy density $V_0$ than is usually assumed, see e.g.\ \cite{Nakayama:2010xf, Buchmuller:2013dja} for recent analyses. In this parameter range both potential gravitino overproduction and the possible formation of cosmic strings at the end of inflation are potentially dangerous and require careful treatment. The investigation of these model-dependent constraints is however beyond the scope of the present paper.\footnote{Cosmic strings with a string tension $\mu \sim \Lambda^2$~\cite{Hindmarsh:2011qj} would be in serious tension with the Planck data~\cite{Ade:2013xla}. This might be avoided by reducing the cosmic string tension through a coupling to the MSSM Higgs fields~\cite{Hindmarsh:2012wh} or by considering GUT groups which, against generic expectation~\cite{Jeannerot:2003qv}, do not produce topological defects, e.g.\ flipped SU(5). For avoiding non-thermal gravitino overproduction, cf.\ e.g.~\cite{Nakayama:2012hy}.}
A further interesting question is the impact of the full two-field dynamics in the complex inflaton plane. Additional fields introduce extra friction, allowing for slow-roll on steeper potentials~\cite{Liddle:1998jc}, and for complicated trajectories the Lyth bound  on the total length of the trajectory can still allow for small total field excursions. A extreme example of the latter point was recently given in~\cite{McDonald:2014oza}. However, even the 
introduction 
of a single $R$-symmetry violating term in $F$-term hybrid inflation can yield non-trivial trajectories, cf.~\cite{Buchmuller:2014epa}.

Finally, a crucial task will be to verify the BICEP2 signal in an independent experiment. Considering the tension with the Planck data and uncertainties involving the modelling of the dust foreground, upcoming data from ABS, ACTPol, EBEX, Planck, POLARBEAR, Spider and SPT will hopefully provide a clearer picture. As Fig.~\ref{fig_sigma_vs_r} demonstrates, a value of the tensor-to-scalar ratio somewhat smaller than the current best-fit value $r = 0.16$ would render this scenario less contrived (but also less testable), while a larger value can hardly be accommodated.

\vspace{1cm}
\subsubsection*{Acknowledgements}

We thank Ido Ben-Dayan, Shaun Hotchkiss, and David Seery for useful discussions. This work has been supported in part by the European Union FP7-ITN INVISIBLES (Marie Curie Action PITAN-GA-2011-289442-INVISIBLES) (VD) and by ERC Advanced Grant 267985 ``Electroweak Symmetry Breaking, Flavour and Dark Matter'' (FB). VS is supported  by the Science and Technology Facilities Council (grant number  ST / J000477 / 1). This project was initiated at the March 2014 workshop on the ``Implications of the 125 GeV Higgs Boson'' at LPSC Grenoble. 

\appendix

\section{A model of dynamical hybrid inflation with explicit {\it R} breaking
\label{app_ddr}}

An example for a model of $F$-term hybrid inflation with explicit $R$-symmetry breaking was constructed by Dimopoulos, Dvali and Rattazzi (DDR) in \cite{ddr}, and will be briefly reviewed and slightly generalized here. We will discuss the limit $\MP\into\infty$ of rigid supersymmetry first. Following \cite{ddr, ddr2} we consider $\SU{N_c=n}$ supersymmetric gauge theory with $N_f=n$ pairs of quarks $Q_I$ and antiquarks $\widetilde Q^I$. This theory is asymptotically free because $b=3N_c-N_f=2 n>0$, and it becomes strongly coupled at a scale $\Lambda$. We add a singlet $S$ and a superpotential 
\be
W=\lambda\,S Q_I\widetilde Q^I \,.
\label{eq_Wtree0}
\ee
This theory has a $\U{1}_R$ symmetry under which $S$ carries $R$-charge 2 and the $Q_I$, $\widetilde Q^I$ are neutral.
For $S\gg\Lambda$ the quarks decouple at the scale $\lambda S$, below which the theory reduces to $\SU{n}$ super-Yang-Mills theory with beta function coefficient $b'=3\,n$ and scale $\Lambda'$. One-loop matching at the scale $\lambda S$ yields
\bea
b\log\frac{\lambda S}{\Lambda}=b'\log\frac{\lambda S}{\Lambda'} \quad
\Rightarrow\;(\Lambda')^3=\lambda\,S\,\Lambda^2\,.
\eea
Gaugino condensation $W_{\rm eff}=(\Lambda')^3$ in the super-Yang-Mills theory thus induces an effective superpotential for $S$, valid at scales $|S|\gg \Lambda/\sqrt{\lambda}$, which is exactly the inflaton superpotential of FHI Eq.~\eqref{WFHIglobal}:
\be
W_{\rm eff}=\lambda\,\Lambda^2\,S\,.
\label{Weff}
\ee
At large $S$ the massive quarks $Q_I$ and $\widetilde Q_I$ generate a logarithmic one-loop correction to the scalar potential. Together with the superpotential Eq.~\eqref{Weff} this results in the inflaton potential of FHI,
\be
V(S)=\lambda^2\Lambda^4+V_{\rm CW}(S)\,,
\ee
where $V_{\rm CW}(S)$ is given by Eq.~\eqref{VCW} with $C=n^2$.

In this model the scale of inflation can be identified with the scale of grand unification when $S$ is coupled to GUT-symmetry breaking such that its vacuum expectation value provides a mass to the GUT-breaking field $\Sigma$ \cite{ddr}. Taking $\Sigma$ to be in the ${\bf 24}$ of $\SU{5}$,\footnote{More realistically one should perhaps consider other groups than $\SU{5}$, since the breaking of $\SU{5}$ to the Standard Model at energies below the scale of inflation may produce magnetic monopoles which cannot be inflated away. 
Also symmetry breaking patterns which produce cosmic strings at the end of inflation are dangerous if the symmetry breaking scale is indeed as high as indicated by the BICEP2 results.
We use $\SU{5}$ here for illustration because our main interest is in the physics of inflation, which largely does not depend on the details of GUT breaking.} the tree-level superpotential \,,
\be\label{Wtree}
W=\lambda\,S Q_I\widetilde Q^I-\frac{\lambda'}{2}\,S\,\tr\Sigma^2+\frac{h}{3}\tr\Sigma^3 \,,
\ee
gives rise to the following effective superpotential at low energies:
\be
W_{\rm eff}=A(\det M-B\widetilde B-\Lambda^{2 n})+\lambda S\tr M+\frac{\lambda'}{2}\,S\,\tr\Sigma^2+\frac{h}{3}\tr\Sigma^3\,.
\ee
Here $M_{IJ}=Q_I\widetilde Q_J$, $B=\epsilon^{I_1\ldots I_{n}}Q_{I_1}\cdots Q_{I_{n}}$, and  $\widetilde B=\epsilon_{I_1\ldots I_{n}}\widetilde Q^{I_1}\cdots\widetilde Q^{I_{n}}$ are composite meson and baryon superfields, and $A$ is a Lagrange multiplier enforcing the quantum deformed moduli space constraint \cite{Seiberg}. This theory has an isolated supersymmetric vacuum at
\be\begin{split}
A=-\sqrt{\frac{n}{15}}\;\frac{\kappa^3 h}{\Lambda^{2n-3}}\,,&\qquad
S=\sqrt{\frac{n}{15}}\;\frac{\kappa h}{\lambda'}\Lambda\,,\\
\Sigma=-\sqrt{\frac{n}{15}}\;\kappa\Lambda\;{\rm diag}(2,\,2,\,2,\,-3,\,-3)\,,&
\qquad M_{IJ}=\Lambda^2\delta_{IJ}\,,\qquad B=\widetilde B=0\,,
\end{split}
\ee
where we have defined
\be
\kappa=\sqrt{\frac{\lambda}{\lambda'}}\,.
\ee
Note that the $F$-flatness condition for $S$ enforces that the $\Sigma$ and $M$ VEVs are proportional. Also note that the last term in Eq.~\eqref{Wtree} explicitly breaks the U(1) $R$-symmetry. There is now no obvious symmetry reason why $S^2$ and $S^3$ terms in $W$ should be absent, but  setting their coefficients to negligibly small values is of course technically natural thanks to the non-renormalization theorem.

As long as the $\SU 5$ gauge coupling is small, the $\SU 5$ dynamics will not significantly affect the dynamically generated superpotential for large $S$ Eq.~\eqref{Weff}. If $|\lambda'|<|\lambda|$, $\Sigma$ is stabilised at zero throughout the inflationary phase and the additional logarithmic contribution from $\Sigma$ to the K\"ahler potential Eq.~\eqref{KFHI} will be subdominant.

For studying the effects of $M_{\rm P}$-suppressed operators on this model, it needs to be embedded into supergravity. In order to end up in a vacuum with (approximately) vanishing cosmological constant after inflation, we add a constant term,
\be
W_0=-\frac{h}{3}\,\langle{\tr\Sigma^3}\rangle = -10 h \left( \frac{n \kappa }{15} \right)^{3/2} \Lambda^3 \,,
\ee
to the superpotential. Thus the DDR model provides an example for a model in which $W_0$ is not given by the gravitino mass after SUSY breaking (in Planck units), but is instead of the order of $M_{\rm GUT}^3$.

\section{Higher-order contributions to the scalar potential
\label{app_coeff}}

The coefficients in the inflaton potential in Eq.~\ref{Vsigma2} can be written in terms of $W_0$ and the $k_{ij}$ coefficients in Eq.~\eqref{KFHI}. Here we show the leading few terms in this expansion, assuming real $k_{ij}$ and $W_0$:
\be\begin{split}
&a_1 = -2\sqrt{2}\,\left(k_{12} +\tilde{w}_0\right) \,,\\
&a_2= \left(\left(8\,k_{12}^2-3\,k_{13}-2\,k_{22}\right)+\left(3\,k_{03}-k_{12}^2\right)\tilde{w}_0-\tilde{w}_0^2\right) \,,\\
&a_3=\frac{1}{\sqrt{2}}\Biggl(6\,k_{23}-16\,k_{12}k_{22}+4\,k_{14}-24\,k_{12}k_{13}-4\,k_{03}+32\, k_{12}^3+2\,k_{12} \\
&\qquad\qquad+\left(2\,k_{22}+2\,k_{13}+12\,k_{03}k_{12}-4\,k_{12}^2-4\,k_{04}+1\right) \, \tilde w_0 +2\,k_{12}\, \tilde w_0^2\Biggr)\,,\\
&a_4=\frac{1}{8}\Biggl(18\,k_{33}+32\,k_{24}-192\,k_{12}k_{23}+64\,k_{12}k_{03}-32\,k_{22}^2-96\,k_{13}k_{22}\\
&\qquad\qquad+384\,k_{22}k_{12}^2+14\,k_{22}+20\,k_{15}+128\,k_{12}k_{14}-72\,k_{13}^2\\
&\qquad\qquad+576\,k_{12}^2k_{13}+16\,k_{13}-20\,k_{04}-512\,k_{12}^4-32\,k_{12}^2-1\\
&\qquad\qquad+\Bigl(28\,k_{23}+18\,k_{13}+12\,k_{22}-48\,k_{12}^2+12\,k_{03}-48\,k_{12}k_{22}\\
&\qquad\qquad\qquad+12\,k_{14}-56\,k_{12}k_{13}-64\,k_{04}k_{12}+64\,k_{12}^3+12\,k_{12}-20\,k_{05}-5\Bigr) \, \tilde w_0\\
&\qquad\qquad+\left(12\,k_{03}k_{12}+6\,k_{22}-4\,k_{04}+8\,k_{13}-2\,k_{12}^2-18\,k_{03}^2+1\right)\, \tilde w_0^2\Biggr)\,,
\label{Vsigma}
\end{split}
\ee
where
\be
\tilde{w}_0=\frac{ W_0}{\lambda\Lambda^2\MP}\ .
\ee
Note that in the main text we are assuming that the vacuum energy during inflation is dominated by the $\lambda^2\Lambda^4\equiv V_0$ term from global SUSY, i.e.~$\tilde{w}_0\ll 1$.

 \providecommand{\href}[2]{#2}\begingroup\raggedright\endgroup

 \end{document}